\documentclass{llncs}
\usepackage{amssymb}
\usepackage{amsmath}
\usepackage{listings}
\usepackage{epstopdf}
\usepackage{graphicx}
\usepackage{subfig}
\global\long\def\Id{\mbox{Id}}
\begin{document}

\title{Efficient Document Indexing Using Pivot Tree}

\author{Gaurav Singh\inst{1} \and Benjamin Piwowarski\inst{2}}
\institute{UPMC, Paris, \\
\email{gaurav.singh.15@ucl.ac.uk}
\and
UPMC, Paris, \\
\email{benjamin.piwowarski@lip6.fr }
}

\maketitle

\begin{abstract}
We present a novel method for efficiently searching top-k neighbors for documents represented in high dimensional space of terms  based on the cosine similarity. Mostly, documents are stored as bag-of-words tf-idf representation. One of the most used ways of computing similarity between a pair of documents is cosine similarity between the vector representations, but cosine similarity is not a metric distance measure as it doesn't follow triangle inequality, therefore most metric searching methods can not be applied directly. We propose an efficient method for indexing documents using a pivot tree that leads to efficient retrieval. We also study the relation between precision and efficiency for the proposed method and compare it with a state of the art in the area of document searching based on inner product. \end{abstract}

\section{Introduction and Related Work}
There are two main areas of research in information retrieval 1.) Search in metric spaces 2.) Search in non-metric spaces. A metric space basically refers to a similarity measure which follows all the metric properties like reflexivity, symmetry, non-negativity and triangle-inequality. All other properties can be achieved by trivial transformations, but triangle inequality is considered most important out of all others, since it is difficult to achieve it using trivial transformations and it can be effectively used in pruning elements. On the other hand, a non-metric space basically refers to a similarity measure that doesn't follow triangle inequality. In such spaces, metric access methods using triangle inequality can not be applied directly. The retrieval approaches in non-metric spaces can be broadly categorized as \textit{embedding} and \textit{classification}.   

A metric space is described by a similarity measure  that follows the properties of reflexivity, nonnegativity, symmetry and most importantly triangle inequality. A number of methods have been developed in the past to search metric spaces\cite{ref:cha,ref:sam,ref:zez}, most metric spaces can be search efficiently using the triangle inequality. \cite{ref:sei,ref:hja} discuss the use of range queries and kNN in metric space. A lot of research has been done in the field of information retrieval from nonmetric spaces, but a lot of similarity measures do not follow triangle inequality. In such non metric spaces, there are broadly two approaches followed, embedding and classification

Embedding basically refers to the conversion of nonmetric space into metric spaces. There are certain embedding methods which perform exact conversion\cite{ref:che} to metric space whereas others do approximate conversion to metric space\cite{ref:ath}. A number of approximate embedding methods have been developed such as \cite{ref:ath} that converts data objects into vector space. They introduce a query sensitive distance function together with the embedding method, in order to give different importance to different embedding dimensions for each query object. TriGen is another method developed to convert nonmetric spaces into metric spaces by using metric preserving and similarity invariant modifiers. But the author himself acknowledges, not all nonmetric measures are suitable to be converted by these methods. In the case of exact embedding methods, the most prominent is LCE\cite{ref:che}, which tried to divide objects into groups and then adds a small local constant to all pairwise distances within a group to make them follow triangle inequality. This method although exact, becomes completely unscalable for large datasets, since it requires the computation of all possible triplets of objects within a cluster, which can be a huge computational cost. A number of other approximate embedding techniques like Fastmap\cite{ref:fal}, Metric Map\cite{ref:wang}, and Sparse Map\cite{ref:hri} exist, but the only exact method with no false dismissals are LCE and CSE\cite{roth2002going}  

\cite{becker2007non} presented a non-metric clustering method based on distances to the so-called fiduciary templates (some selected random objects from the set). The distances to these fiduciary templates form a vector, which is used to decide in which cluster a new object belongs. \cite{ackermann2010clustering} proposed a k-median clustering algorithm for nonmetric functions (specifically, the Kullback-Leibler divergence) that computes $a (1 + \epsilon)-$ approximation of the k-median problem.

Recently \cite{Ram:2012:MIS:2339530.2339677} published maximum inner product based appraoch for querying documents. Their approach is based on creating tigher bounds as the query object traverses down the tree because of reduced number of documents at each new level and therefore a reduced radius. In the proposed method, we project query object on a set of orthogonal pivots as we descend down the pivot tree. We use the previous pivots to construct an orthogonal pivot to all other pivots in the descend path of the query object. We avoid any euclidean addition/subtraction operations that are expensive in high dimensional spaces. The proposed method is based on maximimizing the projection for group of documents on a set of orthogonal projectors. 

\section{Proposed Method}

We observe experimentally that the following relation holds for any given query $q \in \mathbb{R}^{v}$, where v is the vocabulary size, projector $S \in \mathbb{R}^{v \times v}$, document $d \in \mathbb{R}^{v}$ and orthogonal projector  $S^{\bot} \in \mathbb{R}^{v \times v}$


\begin{eqnarray}
q^T d & \le & \|Sq\| \|Sd\|+ \|S^{\bot}q\| \|S^{\bot}d\|\\
 & \le & 1+2\|Sq\| \|Sd\|-\|Sq\|-\|Sd\|
\end{eqnarray}

We can bound the distance between a given document $d$ and query $q$ using the above inequality. We use the above inequality to bound $\|q^Td\|$ for all documents contained in the subtree rooted at node $N_p$, we select a radom pivot $p_{n+1}$ from all such documents.

\subsection{Updating the Projector}
We construct basis $B_n$ for the subspace spanned by the vectors (pivots) $p_1,....,p_n$ in the descend path to the node $N_p$ from the root of the tree.

\[
B_{n}=P_{n}A_{n}\mbox{ with }P_{n}=\left(\begin{array}[t]{ccc}
p_{1} & \ldots & p_{n}\end{array}\right)
\]

Let $p_{n+1}$ be the new vector to be added to the subspace then, we have the new basis $B_{n+1}$: 
\[
B_{n+1}=\left(\begin{array}[t]{cc}
B_{n} & x\end{array}\right)
\]
such that:
\[
x=\frac{y}{\left\Vert y\right\Vert }\mbox{ with }y=\left(\Id-B_{n}B_{n}^{\dagger}\right)p_{n+1}
\]

We can get a projection vector($y$) orthogonal to $B_n$ using the relation:
\[
\left\Vert y\right\Vert ^{2}=\left\Vert p_{n+1}\right\Vert ^{2}-\left\Vert B_{n}B_{n}^{\dagger}p_{n+1}\right\Vert ^{2}=\left\Vert p_{n+1}\right\Vert ^{2}-\left\Vert B_{n}^{\dagger}p_{n+1}\right\Vert ^{2}
\]

Then, denoting $\alpha=\left\Vert y\right\Vert ^{-1}$,
\begin{eqnarray}
B_{n+1} & = & \left(\begin{array}[t]{cc}
P_{n}A_{n} & \alpha\left(\Id-B_{n}B_{n}^{\dagger}\right)p_{n+1}\end{array}\right)\\
 & = & \left(\begin{array}[t]{cc}
P_{n} & p_{n+1}\end{array}\right)\left(\begin{array}[t]{cc}
A_{n} & -\alpha A_{n}A_{n}^{\dagger}P_{n}^{\dagger}p_{n+1}\\
0 & \alpha
\end{array}\right)
\end{eqnarray}

\subsection{Updating the Similarity}
We compute the value of $\left\Vert B_{n+1}^{\dagger}D\right\Vert$ from $\left\Vert B_{n}^{\dagger}D\right\Vert $ for all the documents($D$) contained in the subtree rooted at node $N_p$. Each node of pivot tree contains  $max(\left\Vert B_{n+1}^{\dagger}D\right\Vert^2)$ and $min(\left\Vert B_{n+1}^{\dagger}D\right\Vert^2)$ $\forall D \in D_p$ where $D_p$ is the set of all documents contained in the subtree rooted at node $N_p$.

\begin{eqnarray}
\left\Vert D^{\dagger}B_{n+1}\right\Vert ^{2} & = & \left\Vert \left(\begin{array}[t]{cc}
D^{\dagger}P_{n} & D^{\dagger}p_{n+1}\end{array}\right)\left(\begin{array}[t]{cc}
A_{n} & -\alpha A_{n}A_{n}^{\dagger}P_{n}^{\dagger}p_{n+1}\\
0 & \alpha
\end{array}\right)\right\Vert ^{2}\\
 & = & \left\Vert \left(\begin{array}[t]{cc}
D^{\dagger}P_{n}A_{n} & \alpha D^{\dagger}p_{n+1}-\alpha D^{\dagger}P_{n}A_{n}A_{n}^{\dagger}P_{n}^{\dagger}p_{n+1}\end{array}\right)\right\Vert ^{2}\\
 & = & \left\Vert D^{\dagger}B_{n}\right\Vert ^{2}+\left\Vert \alpha D^{\dagger}p_{n+1}-\alpha D^{\dagger}P_{n}A_{n}A_{n}^{\dagger}P_{n}^{\dagger}p_{n+1}\right\Vert 
\end{eqnarray}

\subsection{Algorithm}
In this section we describe the algorithm we use to construct the pivot tree. We then describe an algorithm to search the pivot tree using a given query.\\\\
\begin{enumerate}

\item Algorithm SelectPivot(Data S)
\begin{lstlisting}[mathescape]
SelectPivot(Data S)
    Pick some random pivots P $\in$ S
    Choose a random pivot p $\in$ P s.t. $argmax_p (\sum \Vert p^TpS_i \Vert^2 ) ~~\forall S_i \in S $    
    return (p)
\end{lstlisting}

\item Algorithm MakeSplit(Data S, Pivot p)
\begin{lstlisting}[mathescape]
MakeSplit(Data S, Pivot p)
    A $\leftarrow \{ s \in S$: $ \Vert D^Tp_{n+1} \Vert ^2 >c\}$ 
    B $\leftarrow S/A$
    return (A,B)
\end{lstlisting}

\item Algorithm UpdateProjections(Data $D_l$, Pivot p, A)
\begin{lstlisting}[mathescape]
UpdateProjections((Data $D$, Pivot p, A)
    $D_i$.Projections $\leftarrow$ update($D_i$,p,A)  $\forall D_i \in D$; # Using eqn. 5

  
\end{lstlisting}

\item Algorithm BuildTree(Data S)
\begin{lstlisting}[mathescape]
BuildTree(Data S)
   Input $\leftarrow$ S
   Output $\leftarrow$ Tree T
   T.S $\leftarrow$ S
   T.min  $\leftarrow$ min(S.Projections)
   T.max $\leftarrow$ max(S.Projections)
   if ($\vert S \vert \leq N_o$)
        return T
   T.p $\leftarrow$ SelectPivot(Data S)
   $D_l$,$D_r$ $\leftarrow$ MakeSplit(T.S, T.p)
   # Using eqn. 4
   T.A $\leftarrow$ UpdateA(pivot p)
   # Using eqn. 4
   T.P $\leftarrow$ UpdateP(pivot p)
   # Using eqn 7.
   $D_l$.Projections $\leftarrow$ UpdateProjections(Data $D_l$, Pivot p, T.A)
   T.left $\leftarrow$ BuildTree(Data $D_l$)
   T.right $\leftarrow$ BuildTree(Data $D_r$)
   return T
\end{lstlisting}

\item Algorithm SearchTree(Query S, Tree T)
\begin{lstlisting}[mathescape]
SearchTree(Query S, Tree T)
   Input $\leftarrow$ Query S, Tree T
   Output $\leftarrow$ Document Set D
   $B_l$ $\leftarrow$ ComputeBound(Tree T.left, Query q) # using eqn 2
   $B_r$ $\leftarrow$ ComputeBound(Tree T.right, Query q) #using eqn 2
   
   #getLast: Returns the element with least similarity with query
   if $(B_l \geq getLast(queue))$
        searchL=True
   if $(B_r \geq getLast(queue))$
        searchR=True    
        
   if(searchL and searchR)
        if ($B_l> B_r$)
            queue $\leftarrow$ SearchTree(Query S, Tree T.left)
        else
           queue $\leftarrow$ SearchTree(Query S, Tree T.right)
    else if (searchL and !searchR)   
        queue $\leftarrow$ SearchTree(Query S, Tree T.left)
    else if (!searchL and searchR)
        queue $\leftarrow$ SearchTree(Query S, Tree T.right)
    else 
        return (queue)
        
   
\end{lstlisting}

\end{enumerate}
\section{Experimentation and Results}
We present in this section experimental results for the proposed method based on the  \textbf{MTA} (Maximized Trace Approach) against state of the art method \textbf{MIP}(Maximum Inner Product) appraoch. The precision versus prunes is drawn for both appraoches by reducing the bound artificially, reduction in bound leads to more prunes, but reduced precision. We can see in Figure 1 that MTA outperforms MIP\cite{Ram:2012:MIS:2339530.2339677} in terms of both ranking (as measured by spearman distance) and precision for different values of prunes.

\begin{figure}
\centering
\includegraphics[scale=0.27]{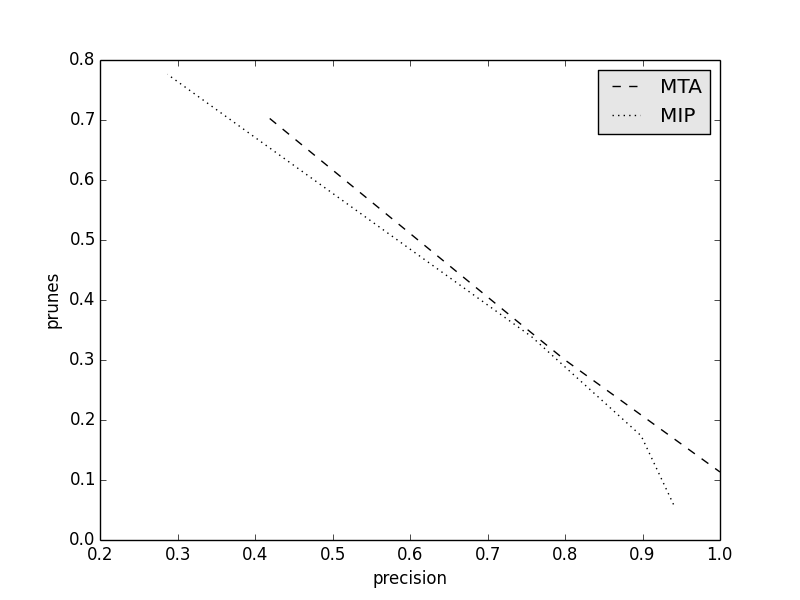}
\includegraphics[scale=0.27]{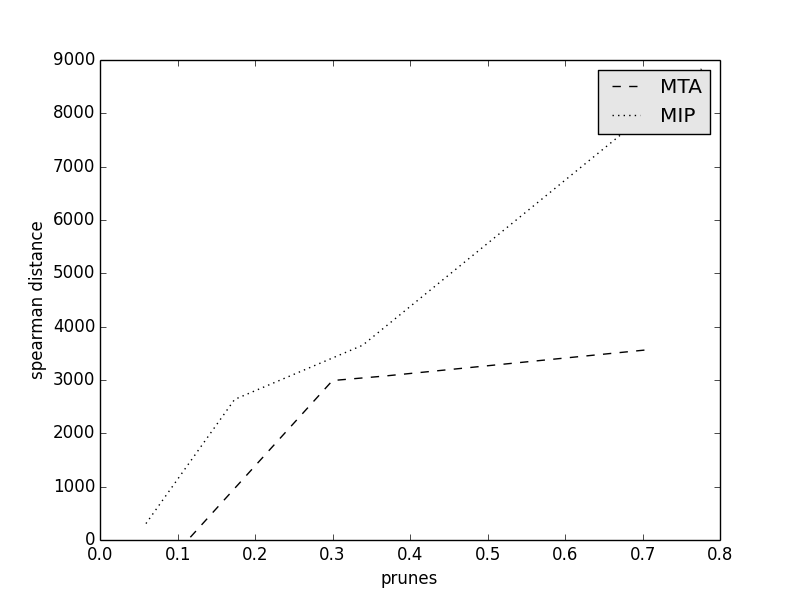}
\caption{The figure in the left presents the prunes against precision. The figure on the right presents ranking performance of the two methods for different number of prunes.}
\label{fig:my_label}
\end{figure}

\bibliographystyle{plain}
\bibliography{ref}
\end{document}